\documentstyle[prb,aps]{revtex}
\newcommand{\beq}{\begin{eqnarray}}
\newcommand{\eeq}{\end{eqnarray}}
\begin{document}
\draft
\twocolumn

\title
{Superconductivity in a two-dimensional Electron Gas}
\author{Philip Phillips, Yi Wan, Ivar Martin, Sergey Knysh, and Denis Dalidovich}
\vspace{.05in}

%
\address
{Loomis Laboratory of Physics\\
University of Illinois at Urbana-Champaign\\
1100 W.Green St., Urbana, IL, 61801-3080}

%
\maketitle


\columnseprule 0pt
\narrowtext

{\bf In a series of recent experiments, Kravchenko and
colleagues\cite{krav,krav1} observed unexpectedly that a
two-dimensional electron gas in zero magnetic field can be a
conductor.  The two-dimensionality was imposed by confining the
electron gas to move laterally at the interface between two
semiconductors.  The observation of a conductor in two dimensions
(2D) is surprising as the conventional theory of metals precludes
the presence of a metallic state at zero temperature in
2D\cite{anderson}.  Nonetheless, there are now several experiments
confirming the existence of the new conducting phase in a dilute
two-dimensional electron gas in zero magnetic
field\cite{popovic,pepper,insup,coleridge}. Here we argue based on
an analysis of the experiments and general theoretical grounds
that this phase is a zero-temperature superconductor with an
inhomogeneous charge density.}

While the specific details differ as the semiconductor that
confines the 2D electron gas is changed, several key similarities
\cite{krav,krav1,popovic,pepper,insup,coleridge} exist among the
reported observations of the transition to the conducting state:
1) the existence of a critical electron or hole density, $n_c$,
above which the conducting phase appears, 2) a characteristic
temperature, typically on the order of half the Fermi temperature,
$T_F$, below which the resistivity on the conducting side
decreases, 3) critical scaling, indicative of a quantum critical
point, in the vicinity of the insulator-conducting phase
transition, 4) non-linear current-voltage (I-V) curves that
exhibit a symmetry around the I-V curve at criticality, and 5)
suppression of the conducting phase by a magnetic field.  In Si
metal-oxide-field-effect-transistors (MOSFET's)
samples\cite{krav}, temperature (T) and electric field (E) scaling
have made it possible to extract the dynamical and correlation
length exponents, $z$ and $\nu$, respectively.  By fitting the
resistivity measurements to functions of the form,
$\rho(T,n)=f_1(|\delta|/T^b)$ and $\rho(E,n)=f_2(|\delta|/E^a)$,
with n, the electron density, $b=1/z\nu$, $a=1/[(z+1)\nu]$, and
$\delta=(n-n_c)/n_c$, Kravchenko and colleagues\cite{krav} found
that $\nu=1.5$ and $z=0.8$.  Analogous scaling occurs in
GaAs\cite{pepper} but only in the most disordered samples.  This
indicates that the underlying transition in the clean system is
first order, whereas in the presence of disorder, it becomes
second order and critical scaling applies.

Three postulates anchor the conventional theory of metals\cite{anderson}:
1) Fermi liquid theory accurately
describes the low-temperature physics of conventional clean metals, 2) the classical
and quantum corrections to the conductivity are additive\cite{ln} as the system
size increases, and 3) the
logarithmic derivative ($\beta$) of the dimensionless conductance with respect to
the system size is a continuous
monotonic single-valued function as the strength of the disorder is
increased from weak to strong. From (2) and (3) it follows\cite{anderson,ln} that $\beta$ is
always negative in two dimensions.  Hence, as the system
size increases, the conductance decreases and
insulating behavior necessarily obtains in two dimensions.
The general applicability of the scaling analysis rests firmly on the extensive numerical and experimental
work that have confirmed its central predictions and assumptions\cite{lr}.

Strictly speaking, however, Fermi liquid behaviour obtains only for $T\ll T_F$.
Hence, the onset of a conducting
phase at $T\approx T_F$ does not necessarily pose a paradox provided that the
resistivity turns upwards at sufficiently low temperatures.  However,
such behaviour is inconsistent with
the experimental observations for two primary reasons. First,
there is no experimental indication that the resistivity turns upwards at low
temperatures.
Secondly, the critical scaling observed\cite{krav} signifies
that the conductor-insulator transition is a bona-fide quantum phase transition.

The conventional theory of metals forces us then to come to grips
with the new conducting state in 2D by abandoning one or all of
the postulates of the standard view.  In the context of the
experiments\cite{krav,krav1,popovic,pepper,insup,coleridge}, the
Coulomb interaction $V_{\rm ee}$ exceeds the Fermi energy,
$\epsilon_F$, by at least an order of magnitude.  For long-range
interactions, non-Fermi liquid behaviour is likely to obtain when
$V_{\rm ee}\gg\epsilon_F$.  It is for this reason and the
localization constraint placed on Fermi liquids in 2D by the
conventional theory that the experiments force us to consider
non-Fermi liquid scenarios only. In this vein, Chakravarty and
co-workers\cite{chak} proposed a Luttinger liquid state to explain
the experimental observations. However, such a state has yet to be
proven to exist except in 1-dimension.  In fact in 2D, a $T=0$
superconductor is the only conducting non-Fermi liquid state
proven to exist in the presence of disorder\cite{dirty} and zero
magnetic field.  Hence, rather than speculate as to the existence
of some yet-unproven non-Fermi liquid state in 2D, we take the
conservative approach and propose that superconducting
fluctuations mediate the new conducting state in 2D.

We support this proposal with three subsidiary arguments:
1) the generic features\cite{pw} of the conducting
transition resemble those of known\cite{ali} insulator-superconductor phase transitions,
2) the magnetoresistance
measurements\cite{krav,pepper}  provide evidence for the existence of a critical magnetic
field above which the conducting phase is destroyed, and
3) the conducting transition lies in close proximity
to an incipient electron crystal state in which strong charge
retardation effects can lead to Cooper pair formation.
Further, we draw a parallel between the 2D electron gas experiments
and those on inhomogeneous superconductors in which
an apparent saturation of the resistivity has been observed
at low temperatures\cite{jaeger,imry}.  We argue that the enhanced role of classical
fluctuations of the superconducting phase\cite{emery} in 2D can
significantly suppress the temperature at which the resistivity vanishes.

In support of the superconducting scenario are numerous transport
measurements on the electron and hole systems.  The value of the
dynamical exponent\cite{ali}, $z=1$, the non-linearity in the I-V
curves as well as the observed temperature and electric field
scaling of the resistivity are all consistent with what is
observed in known insulator-superconductor transitions (IST's). We
now analyze the experimental data on the magneto-resistance and
show that they offer evidence for a critical parallel magnetic
field, consistent with singlet pairing. Although we limit our
discussion to a parallel field ($H_{||}$), Kravchenko and
colleagues\cite{krav1} showed that the response of the conducting
phase to a magnetic field is independent of the direction of the
field, indicating that it is a spin effect that destroys the
conducting phase. Consider first the magneto-resistance
measurements in GaAs\cite{pepper}. Measurements of the
conductivity as a function of the hole density in GaAs in the
temperature interval [1.4K,0.3K] reveal that even in the presence
of a magnetic field, the conductivity curves cross at a unique
value of the hole density.  The single crossing point signifies
that the conductivity is temperature independent at a particular
density. Hence, this density demarcates the transition between the
conducting and insulating phases. If the conducting phase were
destroyed by an arbitrarily small magnetic field, such a crossing
point would not occur at finite field. However, the experiments
show that even for fields as high as $1T$, a unique crossing point
exists.  By $H_{||}=3.0T$, the unique crossing point vanishes,
indicating that there is a threshold parallel magnetic field above
which the conducting phase is extinguished.  The critical field
is, however, density dependent.  From the crossing point in finite
field, we conclude that for $\delta=0.02$ and $\delta=0.03$, the
critical fields are $H^c_{||}=0.5T$ and $H^c_{||}=1.0T$,
respectively.  For GaAs, the Zeeman energies at such field
strengths correspond to an energy at least an order of magnitude
smaller than the Fermi temperature.  Hence, the conducting phase
in GaAs is characterized by an internal energy scale distinct from
$\epsilon_F$.

What about Si MOSFET's? Based on the measurements of $\rho(H,E)$,
Kravchenko, et. al.\cite{krav} concluded that the conducting phase
is destroyed for an arbitrarily small parallel magnetic field,
indicating a vanishing of the critical field. This conclusion
poses a distinct problem if the conducting transition in Si
MOSFET's and GaAs is assumed to be driven by the same physics.
That is, either both or neither should display a critical field,
$H_c$. We point out, however, that the observation of a critical
field in GaAs was based on an analysis of $\rho(H,T)$, not
$\rho(H,E)$. If this conclusion is correct, $\rho(H,T)$ should
also confirm this result.  Hence, we investigate precisely what is
contained in $\rho(H,T)$ for Si MOSFET's.  The raw data for Si
MOSFET's indicates that for $H_{||}\approx 9kOe$ the slope of
$\rho(T)$ changes sign\cite{krav1}.  Hence, from the raw data,
there is no indication at least within the temperature regime
studied that an arbitrarily small magnetic field suppresses the
conducting phase. As a consequence, we analyzed $\rho(H,T)$ for a
Si MOSFET (D. Simonian, S.  Kravchenko, and M. Sarachik, personal
communication) with a scaling function of the form,
$f(|H_{||}-H_c|/T^{1/\alpha})$. In field-tuned 2D
IST's\cite{ali,goldman}, the resistivity scales on either side of
$H_c$ as a universal function\cite{goldman} of
$|H-H_c|/T^{1/z_B\nu_B}$, where $z_B$ and $\nu_B$ are the
dynamical and correlation length exponents, respectively in a
magnetic field. If a critical field exists, then above and below
$H_c$ the experimental values for the resistivity should collapse
onto two distinct branches. Such collapse is shown in Fig.
(\ref{scaling}) above and below a critical field of $H_c=9.5kOe$
with $\alpha=0.6\pm 0.1$. Here again, the critical field
corresponds to an energy scale that is an order of magnitude
smaller than the Fermi energy. It is interesting to note that
experiments on the IST in bismuth films (N. Markovic,
Christiansen, and A. M. Goldman, personal communication) show that
$z_B\nu_B=0.7\pm 0.2$, which is remarkably close to the value,
$\alpha=0.6\pm 0.1$, obtained in the scaling plot in Fig.
(\ref{scaling}).

The stiffness in the conducting phase to a parallel magnetic field
signifies that the ground state is a singlet. While spin glass and
antiferromagnetic order are also consistent with a singlet ground
state, such phases are insulating in 2D. If we entertain the
possibility of other non-Fermi liquid states, it is unclear how
such a state will differ from a superconducting one, as the
experiments dictate that such a state must also have a singlet
energy gap and conduct in the presence of disorder.

In terms of the dimensionless measure of the density,
$r_s=1/\sqrt{\pi\rho }a_0^\ast$, $a_0^\ast$ the effective Bohr
radius, the conducting transition for Si MOSFET's\cite{krav} and
GaAs\cite{pepper,insup} occurs at $r_s\approx 10$ and $r_s\approx
18$, respectively.  For such dilute systems, $r_s\gg 1$, it is not
known definitively what type of microscopic order obtains in the
ground state.  However, as a Fermi liquid description is valid for
dense systems, $r_s\alt 1$, perturbation theory\cite{lee,abrahams}
will necessarily fail to describe the experimental
observations\cite{popovic,pudalov}. Monte Carlo
simulations\cite{tc} reveal that in a clean 2D electron gas, a
Wigner crystal is the ground state for $r_s>37$.  A Wigner crystal
is an electron crystalline phase in which the electrons minimize
their repulsive potential energy and form an ordered array. For
clean 2D systems, the electrons arrange themselves in a triangular
lattice. The simulations also indicate that in the presence of
disorder\cite{eliot,ct} or random pinning centers, an incipient
Wigner crystal phase exhibiting quasi-long range order still
forms. The melting density for such a phase is shifted to higher
values, typically $r_s\approx 10$ (though the amount and type of
disorder might change this value) because disorder stabilizes the
solid relative to the liquid phase. As disorder is undoubtedly
present in the experimental systems and the conducting transition
occurs for $r_s\approx 10$, it makes sense to think about the
transition to the conducting state as a transition from an
insulating incipient Wigner crystal phase.  Experimentally, the
density dependence of the threshold electric field\cite{krav2}
required to initiate transport on the insulating side, $E_t\propto
\delta^{1.5}$, is consistent with the prediction\cite{cqt} for
quantum tunneling in an incipient Wigner crystal. Hence,
experimental evidence also corroborates the proximity of the
conducting transition to an insulating phase with quasi-long range
order.

The proximity of the conducting phase to a quasi-crystalline
dilute electron phase makes this problem quite analogous
to high temperature superconductivity (high $T_c$).
While this phenomenon is currently unexplained, experimentally
it is clear that superconductivity in the copper oxides
obtains as doped holes disrupt
the
perfect antiferromagnetic order in the Mott insulating phase.
The correspondence with high $T_c$ is even more striking as
such systems are quasi-2D with $r_s\sim 10$ for the holes.  In both systems, destruction
of the long-range or quasi-long range spin or charge order in the
insulating phase is accompanied by a charge or spin retardation
effect which attempts to preserve the memory of the correlations in
the insulating phase.
As an incipient Wigner crystal has at best frustrated antiferromagnetic\cite{tc}
spin order, the charge retardation effect is expected to dominate\cite{takada}.
We illustrate its role as follows.

Consider an electron crystalline or quasi-crystalline phase in
which the electrons are locked into `home' positions by the
electron interaction.  Such a crystalline phase is stable when the
zero-point energy $\hbar\omega_0$ much exceeds the kinetic energy,
$\epsilon_F$, of an electron in each unit cell (or Wigner-Seitz
cell in the context of a Wigner crystal) of edge length $r_s$.  A
distinct feature of an electron crystal or quasi-crystalline phase
is the dominant role played by the correlations. The correlations
manifest themselves in the form of a correlation hole which is
centered at the average position of each electron in a unit cell.
Increasing the electron density leads to an increase in the
electron kinetic energy and an eventual melting of a Wigner
crystal. In the melted phase, correlation holes still form around
electrons.  However, because they are massive, relative to a
single electron, their response is delayed. The retardation
timescale is set by the inverse of the plasma frequency. Within
this timescale, a correlation hole lags behind its associated
electron and hence appears positively charged to another electron.
As a consequence, the partially-vacated correlation hole can
attract another electron.  In so doing, the correlation hole
mediates a dynamic attraction between electrons and the subsequent
onset of Cooper pair formation.  Because plasmons are ungapped in
2D, the retardation effect is strongest for the low frequency
plasmons.  We anticipate that it is from such plasmons that the
dominant electron attraction arises.

For a 3D electron gas, Takada\cite{takada,ren} has made the
analogous observation regarding the proximity of superconductivity
to the melting of a Wigner crystal. The work of Kelly and
Hanke\cite{hanke} and Ren and Zhang\cite{ren} also relates to this
proximity. More recently, Belitz and Kirkpatrick\cite{bk} have
proposed a similar charge polarization mechanism in which the
long-time tail of the charge density correlation function, which
appears in the presence of disorder, assists pairing. Also, in the
Monte Carlo simulations\cite{tc} on a clean Wigner crystal, a
precipitous drop of the spin susceptibility is observed in the
melted phase, a pre-requisite for singlet superconductivity. In
addition, capacitance charging experiments on GaAs quantum
dots\cite{zhitnev}, in which a 2D electron gas is constructed one
electron at a time from the very first electron, have reported the
occurrence of pair electron charging events in the density range
$2.8<r_s<9.6$.  As pair-tunneling events correspond to two
electrons charging the same quantum state on the dot, such states
form only if an electronic attraction screens the Coulomb
repulsion between the two electrons. In so far as the electrons in
a quantum dot realistically model a dilute ($r_s>3$) 2D electron
gas, it is certainly reasonable to suspect that the electron
attraction persists in Si MOSFET's as well.

In 2D superconductors, the resistivity vanishes not at the
mean-field temperature, $T_c$, at which the pair amplitude is
established but at a lower temperature, $T_{\theta}$, where global
phase coherence obtains. That is, rather than defining the
superconducting transition temperature, $T_c$ determines the
characteristic temperature at which the resistivity initially
decreases. Both disorder\cite{beasley} and low superfluid
densities\cite{emery} can significantly lower $T_{\theta}$
relative to $T_c$ in 2D systems.  In fact, Emery and
Kivelson\cite{emery} have argued that a key feature which
suppresses $T_{\theta}$ in high $T_c$ materials is their
notoriously low superfluid density.  In the electron systems of
interest, disorder is present (leading to inhomogeneous electron
density) and the superfluid density is expected to be small as the
electron gas is dilute.  As a consequence, we anticipate that in
Si MOSFETS's and GaAs, $T_{\theta}$, will be greatly suppressed
relative to the mean-field $T_c$.  In the case of
GaAs,\cite{insup,pepper} exponential drop of the resistivity on
the conducting side is accompanied by a plateau below some
characteristic temperature.  This latter experimental trend is
identical to the behaviour observed by Goldman and
collaborators\cite{jaeger} and Imry and Strongin\cite{imry} in
inhomogeneous 2D superconductors. The strong similarity between
the inhomogeneous systems and the GaAs data suggests that the
phase-locked zero-resistance state occurs at a temperature
significantly lower than that probed experimentally. Nonetheless,
the non-linear I-V characteristics\cite{krav} observed in the Si
MOSFET's is consistent with a 2D superconducting state on the
brink of global phase coherence.  However, as long as
$T>T_\theta$, pair fluctuations are quenched in a magnetic field
only when the Zeeman and pairing energies are comparable.

In the dense limit, correlation and retardation effects are
negligible and consequently, Cooper pair formation ceases.  The
anticipated termination of the electron attraction at small $r_s$
suggests the schematic phase diagram depicted in Fig. (\ref{pd}).
Beyond the upper density at which superconducting pair
fluctuations cease, the electron liquid is most likely insulating.
Hence, a direct transition from an insulator to a metal in 2D by
changing the density seems unlikely.   In GaAs samples, a
re-entrant (M. Y. Simmons, personal communication) insulating
phase at $r_s\approx 8$ was observed consistent with the analysis
here. Further experiments are needed on the MOSFET's to see the
re-entrant insulating phase.

In closing, because Fermi iquids are localized in 2D, the new
conducting state must be some type of non-Fermi liquid. As $T=0$
superconductivity is the only conducting, disordered non-Fermi
liquid proven to exist in 2D, this must be viewed as the leading
candidate to explain the experimental observations.  The existence
of a critical parallel magnetic field is a clear indicator of a
singlet energy gap in the conducting phase in 2D.  However,
further experiments are needed to probe the low temperature
physics as well as the magneto-resistance in Si MOSFET's. In
addition to disorder and a suppressed superfluid density, the
inhomogeneities introduced in a dilute electron gas ($r_s>3$) as a
result of the negative compressibility\cite{ceperley} (an
instability relative to a uniform charge density) are expected to
enhance phase fluctuations as well.  Consequently, we anticipate
that the transition to the phase-locked state will occur at a
temperature significantly below the mean-field $T_c$.

\begin{figure}
\caption{Scaling curve for the magnetoresistance obtained from the experimental
data of Simonian,
Kravchenko, and Sarachik illustrating clearly the existence
of a critical field at $9.5kOe$ with $\alpha=0.6$.}
\label{scaling}
\end{figure}
\begin{figure}
\caption{Schematic mean-field phase diagram for a 2D disordered electron system.
DWC=Disordered Wigner crystal and SC=superconductor.  The black circle at
$n_c$ is determined from the experimental data in Ref. 1}
\label{pd}
\end{figure}
\acknowledgements
We thank Bob Laughlin for his
most generous, copious,
and characteristically level-headed remarks regarding the presentation.
We also acknowledge useful conversations with
E. Fradkin, N. Markovic, A. Goldman,
A. Yazdani, A. Castro-Neto, Tony Leggett, D. Ceperley, T. Giamarchi,
P. Parris, and S. Wan. We also the ACS Petroleum
Research Fund and the NSF grant No. DMR94-96134.
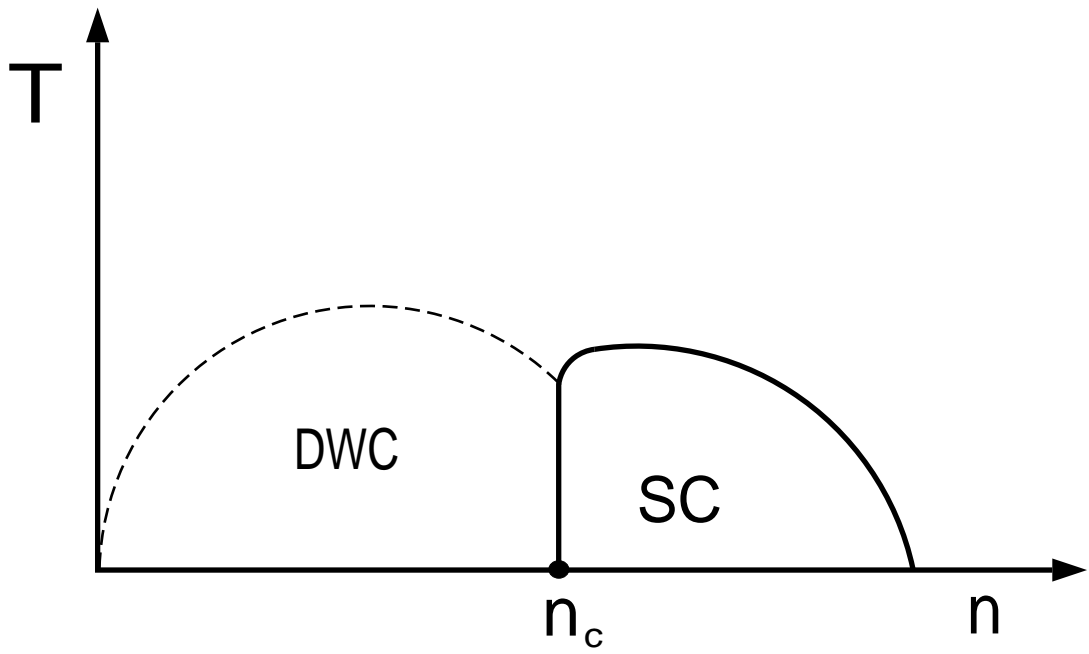
\end{document}